\documentstyle[sprocl]{article}

\def\be{\begin{equation}}
\def\ee{\end{equation}}
\def\bea{\begin{eqnarray}}
\def\eea{\end{eqnarray}}

\begin{document}

\title{SOME PROPERTIES OF TYPE I${}^{\prime}$ STRING THEORY}

\author{ JOHN H. SCHWARZ}

\address{California Institute of Technology, Pasadena, CA 91125, USA}

%

\maketitle
\abstracts{The T-dual formulation of Type I superstring theory, sometimes called
Type I${}^{\prime}$ theory, has a number of interesting features. Here we review some of them
including the role of D0-branes and D8-branes in controlling possible gauge
symmetry enhancement.}

\section{Introduction}

I am pleased to contribute to this volume in memory of Yuri Golfand.  His name
will be remembered by future generations of physicists for his 1971 paper with
Likhtman,\cite{golfand71}
which introduced the four-dimensional super-Poincar\'e algebra for
the first time.  Recognizing that such a symmetry algebra is a consistent
mathematical possibility was certainly a remarkable achievement.  It is a
curious coincidence that this paper appeared within a few days of Pierre
Ramond's paper on fermionic strings.\cite{ramond71}  
Communications were not so good in those
days, and the Golfand--Likhtman work was not generally known (at least in the
West) for several years.  As a result, its influence in driving the development
of supersymmetry was not as great as it should have been.  In fact,
supersymmetric theories in two dimensions were developed to describe the
world-sheet theory of RNS strings,\cite{gervais71}
and this motivated Wess and Zumino to seek
four-dimensional analogs.\cite{zumino74}  
Only years later did we understand that RNS strings,
properly interpreted, have local 10-dimensional spacetime supersymmetry.\cite{gliozzi77}

The version of the theory that received the most attention prior to 1985
was the one containing both open and closed strings, which Mike Green
and I called the type I theory, since it has one ten-dimensional supersymmetry.
In 1984 we showed that this theory is inconsistent (due to gauge anomalies)
unless the gauge group is chosen to be SO(32).\cite{green84}
Then the anomalies cancel, and
consistency is achieved.  In this manuscript, I propose to review some of the
interesting features that appear when one of the spatial dimensions is chosen
to be a circle.  In this case an alternative $T$ dual
description, known as type I${}^{\prime}$, is available.  This description gives a
different viewpoint for understanding various phenomena, such as gauge symmetry
enhancement.  The material presented here is not new, though it may be
organized somewhat differently than has been done before.

\section{T Duality}

Let $X^\mu (\sigma,\tau)$ denote the embedding functions of a closed string
world-sheet in ten-dimensional spacetime.  In the case of a trivial flat
geometry, the world sheet field equations are simple two-dimensional wave
equations.  Suppose that one of the nine spatial dimensions, $X^9$ say, is
circular with radius $R$.  Denoting $X^9$ by $X$ for simplicity, the general
solution of the wave equation is
\begin{equation}
\label{Xform}
X = mR\sigma + \frac{n}{R} \tau + {\rm ~periodic~terms}.
\end{equation}
The parameter $\sigma$ labels points along the string and is chosen to have
periodicity $2\pi$.  Thus $m$ is an integer, called the winding number, which
is the number of times the string wraps the spatial circle.  The parameter $\tau$
is world-sheet time, and correspondingly $p = n/R$ is the momentum
along the circle.  Single-valuedness of $e^{ipX}$ requires that $n$ is an
integer, called the Kaluza--Klein excitation number.

The general solution of the $2d$ wave equation consists of arbitrary
left-moving and right-moving pieces
\begin{equation}
X(\sigma, \tau) = X_L (\sigma + \tau) + X_R (\sigma-\tau).
\end{equation}
In the particular case described above we have
\begin{eqnarray}
X_L &=& \frac{1}{2} \left(mR + \frac{n}{R}\right) (\sigma + \tau) + \ldots
\nonumber\\
X_R &=& \frac{1}{2} \left(mR - \frac{n}{R}\right) (\sigma - \tau) + \ldots .
\end{eqnarray}
T duality is the world-sheet field transformation $X_R \rightarrow - X_R, X_L
\rightarrow X_L$ (or vice versa) together with corresponding transformations of
world-sheet fermi fields.  There are two issues to consider: the transformation
of the world-sheet action and the transformation of the space-time geometry.
The world-sheet action may or may not be invariant
under T duality, depending on the theory, but
the classical description of the
spacetime geometry is always radically changed.  Let us examine that first:
\begin{equation}
X = X_L + X_R \rightarrow X_L - X_R = \frac{n}{R} \sigma + mR\tau + \ldots .
\end{equation}
Comparing with eq.(\ref{Xform}), we see that this describes a closed string
on a circle of radius $1/R$ with winding number $n$ and Kaluza--Klein
excitation number $m$.  Thus we learn the rule that under T duality $R
\rightarrow 1/R$ and $m \leftrightarrow n$.\cite{giveon94} 
In the case of type I or type II
superstrings, world-sheet supersymmetry requires that $\psi_R^9 \rightarrow -
\psi_R^9$ at the same time.  This has the consequence for type II theories of
interchanging the IIA theory (for which space-time spinors associated with
left-movers and right-movers have opposite chirality) and the IIB theory (for
which they have the same chirality).  Thus T duality is not a symmetry in this
case --- rather it amounts to the equivalence of the IIA theory compactified on
a circle with radius $R$ and the IIB theory on a circle with radius $1/R$.  If
we compactified on a torus instead, and performed T duality transformations
along two of the cycles, then this would take IIA to IIA or IIB to IIB and
would therefore be a symmetry.

In recent years, D-branes have played a central role in our developing
understanding of string theory.\cite{polchinski95} 
These are dynamical objects, which can be
regarded as nonperturbative excitations of the theory. They have the property
that open strings can end on them.  When they have $p$ spatial dimensions they are
called D$p$-branes.  If a D$p$-brane is a flat hypersurface, the coordinates
can be chosen so that it fills the directions $X^m, \  m = 0,1, \ldots, p$ and has a
specified position in the remaining ``transverse'' dimensions $X^i = d^i$ where $i = p
+ 1, \ldots, 9$.  An open string ending on such a D-brane is required to
satisfy Neumann boundary conditions in tangential directions
\begin{equation}
\partial_\sigma X^M|_{\sigma = 0} = 0 \qquad m = 0, 1, \ldots , p,
\end{equation}
and Dirichlet boundary conditions in the transverse directions
\begin{equation}
X^i = d^i \qquad i = p + 1, \ldots, 9.
\end{equation}

A remarkable fact, which is easy to verify, is that the T duality
transformation $X_R \rightarrow - X_R$ interchanges Dirichlet and Neumann
boundary conditions.  This implies that an ``unwrapped'' D$p$-brane, which is
localized on the circle, is mapped by T duality into a D$(p + 1)$-brane that is
wrapped on the dual circle.  This rule meshes nicely with the fact that the IIA
theory has stable (BPS) D$p$-branes for even values of $p$ and the IIB theory has
stable D$p$-branes for odd values of $p$.  An obvious question that arises is
how the wrapped D-brane encodes the position along the circle of the original
unwrapped D-brane.  The answer is that a type II D-brane has a U(1) gauge field
$A$ in its world volume, and as a result a wrapped D-brane has an associated
Wilson line $e^{i\oint A}$.  This gives the dual description of position on the
circle.

\section{Type I Superstrings}

Type IIB superstrings have a world-sheet parity symmetry, denoted $\Omega$.
This $Z_2$ symmetry amounts to interchanging the left- and right-moving modes
on the world sheet:  $X_L^\mu \leftrightarrow X_R^\mu, \, \psi_L^\mu
\leftrightarrow \psi_R^\mu$.  This is a symmetry of IIB and not of IIA, because
only in the IIB case do the left and right-moving fermions carry the same
space-time chirality.  When one gauges this $Z_2$ symmetry, the type I theory results.\cite{sagnotti87}  
The projection operator $\frac{1}{2} (1 + \Omega)$ retains the
left-right symmetric parts of physical states, which implies that the resulting
type I closed strings are unoriented.  In addition, it is necessary to add a
twisted sector --- the type I open strings.  These are strings whose ends are
associated to the fixed points of $\sigma \rightarrow 2\pi - \sigma$, which are
at $\sigma = 0$ and $\sigma = \pi$.  These strings must also respect the
$\Omega$ symmetry, so they are also unoriented.  The type I theory has half as
much supersymmetry as type IIB (16 conserved supercharges instead of 32 ---
corresponding to a single Majorana--Weyl spinor).  This supersymmetry
corresponds to the diagonal sum of the $L$ and $R$ supersymmetries of the IIB
theory.

This ``orientifold'' construction of the type I theory has the entire 10d
spacetime as a fixed point set, since $\Omega$ does not act on $x^\mu$.
Correspondingly a spacetime-filling orientifold plane (an O9-plane) results.
This orientifold plane turns out to carry $-32$ units of $RR$ charge, which must
be cancelled by adding 32 D9-planes.  Rather than proving this, we can make it
plausible by recalling that $n$ type I D9-planes carry an SO($n$) gauge group.
Moreover, we know that the total charge must be cancelled and that SO(32) is
the only orthogonal group allowed by anomaly cancellation requirements.
Correspondingly, these are the unique choices allowed by tadpole cancellation.
As a remark on notation, let me point out that instead of speaking of 32
D$9$-branes, we could equivalently speak of 16 D$9$-branes and their mirror
images.  This distinction is simply one of conventions.  The important point is
that when $n$ type I D9-branes and their $n$ mirror images coincide with an O9-plane,
the resulting system has an unbroken SO(2n) gauge symmetry.

\section{The Type I$^\prime$ Theory}

We now wish to examine the T-dual description of the type I theory on a
spacetime of the form $R^9 \times S^1$, where the circle has radius $R$.  We
have seen that IIB is T dual to IIA and that type I is an orientifold
projection of IIB.  Therefore, one should not be surprised to learn that the
result is a certain orientifold projection of type IIA compactified on the dual
circle $\tilde{S}^1$ of radius $R' = 1/R$.  The resulting T dual version of type I has been
named type IA and Type I$^\prime$ by various authors.  We shall adopt the
latter usage here.

We saw that T duality for a type II theory compactified on a circle corresponds
to the world-sheet symmetry $X_R \rightarrow - X_R, \psi_R \rightarrow -
\psi_R$, for the component of $X$ and $\psi$ along the circle.  This implies
that $X = X_L + X_R \rightarrow X' = X_L - X_R$.  In the case of type II
theories, we saw that $X'$ describes a dual circle $\tilde{S}^1$ of radius $R' =
1/R$.  In the type I theory we gauge world-sheet parity $\Omega$, which
corresponds to $X_L \leftrightarrow X_R$.  Evidently, in the T dual formulation
this corresponds to $X' \rightarrow - X'$.  Therefore this gauging gives an
orbifold projection of the dual circle:  $\tilde{S}^1 /Z_2$.  More precisely the
$Z_2$ action is an orientifold projection that combines $X' \rightarrow - X'$
with $\Omega$.  This makes sense because $\Omega$ above is not a symmetry of
the IIA theory, since left-moving and right-moving fermions have opposite
chirality.  However, the simultaneous spatial reflection $X' \rightarrow - X'$
compensates for this mismatch.

The orbifold $\tilde{S}^1 /Z_2$ describes half of a circle.  In other words, it is
the interval $0 \leq X' \leq \pi R'$.  The other half of the circle should be
regarded as also present, however, as a mirror image that is also $\Omega$
reflected.  Altogether the statement of T duality is the equivalence of the
compactified IIB orientifold $(R^9 \times S^1)/\Omega$ with the type IIA
orientifold $(R^9 \times S^1)/ \Omega \cdot {\mathcal I}_1$.  The symbol
${\mathcal I}_1$ represents the reflection $X' \rightarrow - X'$.

The fixed-point set in the type I${}^{\prime}$ construction consists of a pair of orientifold
8-planes located at $X' = 0$ and $X' = \pi R'$.  Each of these carries $-16$ units
of $RR$ charge.  Consistency of the type I${}^{\prime}$ theory requires adding 32
D$8$-branes.  Of these, 16 reside in the interval $0 \leq X' \leq \pi R'$ and
16 are their mirror images located in the interval $\pi R' \leq X' \leq 2\pi
R'$.  Clearly, these D$8$-branes are the T duals of the D$9$-branes of the
type I description.

The positions of the D$8$-branes along the interval are determined in the type
I description by Wilson lines in the Cartan subalgebra of SO(32).  Since this
group has rank 16, its Cartan subalgebra has 16 generators.  Let $A^I$ denote
the component of the corresponding 16 gauge fields along the circular
direction.  These correspond to compact U(1)'s, so their values are
characterized by angles $\theta_I$.  These determine the dual positions of the
D$8$-branes to be
\begin{equation}
X'_I = \theta_I R', \quad I = 1,2, \ldots, 16.
\end{equation}

The SO(32) symmetry group is broken by the Wilson lines to the subgroup that
commutes with the Wilson line matrix.  In terms of the type I${}^{\prime}$ description
this gives the following rules:
\begin{itemize}
\item When $n$ D$8$-branes coincide in the interior of the interval, this
corresponds to an unbroken U($n$) gauge group.
\item When $n$ D$8$-branes coincide with an O8-plane they give an unbroken
SO($2n$) gauge group.
\end{itemize}
In both cases the gauge bosons arise as zero modes of $8-8$ open strings.  In
the second case the mirror-image D$8$-branes also contribute.  As we will
explain later, this is not the whole story.  Further symmetry enhancement can
arise in other ways.

The case of trivial Wilson line (all $A^I = 0$) corresponds to having all 16
D$8$-branes (and their mirror images) coincide with one of the D$8$-branes.
This gives SO(32) gauge symmetry, of course.  In addition there are two U(1)
factors.  The corresponding gauge fields arise as components of the 10d metric
and B field: $g_{\mu 9}$ and $B_{\mu 9}$.  One combination of these belongs to
the 9d supergravity multiplet, whereas the other combination belongs to a 9d
vector supermultiplet.

Somewhat more generally,  consider the Wilson line
\begin{equation}
\left(\begin{array}{cc}
I_{16 + 2N} & 0\\
0 & I_{16 - 2N} \end{array} \right).
\end{equation}
This corresponds to having $8 + N$ D$8$-branes coincide with the O8-plane at
$X' = 0$ and $8-N$ D$8$-branes with the O8-plane at $X' = \pi R'$.  Generically
this gives rise to the gauge symmetry
\begin{equation}
SO(16 + 2N) \times SO(16 - 2N) \times U(1)^2.
\end{equation}
However, from the S-dual heterotic description of the type I theory, one knows
that for a particular value of the radius further symmetry enhancement is
possible.  Specifically, for heterotic radius $R_H^2 = N/8$ one finds the gauge
symmetry enhancement
\begin{equation}
SO(16 - 2N) \times U(1) \rightarrow E_{9 - N}.
\end{equation}
This radius, converted to type I metric, corresponds to $R^2 = gN/8$.  This
symmetry enhancement will be explained from a type I$^\prime$ viewpoint later.
There are other interesting extended symmetries such as SU(18) and SO(34),
which might also be understood from a type I$^\prime$ viewpoint, but
will not be considered here.

\section{D0-Branes}

The type I$^\prime$ theory is constructed as a type IIA orientifold.  As such, its
bulk physics --- away from the orientifold planes --- is essentially that of
the type IIA theory.  More precisely, there are number of distinct type IIA
vacua distinguished by the difference in the number of D$8$-branes to the left
and the right.  When these numbers match, one has the ordinary IIA vacuum.
When they don't one has a ``massive'' IIA vacuum of the kind first considered
by Romans.\cite{romans86}  In any case, the ordinary IIA vacuum admits various
even-dimensional D-branes.  Here I wish to focus on D$0$-branes.  Later we will
discuss what happens to them when they cross a D$8$-brane and enter a region
with a different IIA vacuum.

D$0$-branes of the type I$^\prime$ theory correspond to type I D-strings that
wrap the compactification circle.  The Wilson line on the D-strings controls
the positions of the dual D$0$-branes.  A collection of $n$ coincident type 1
D-strings has an O($n$) world-volume gauge symmetry.  Unlike the case of
D$9$-branes the reflection element is included, so that the group really is
O($n$) and not SO($n$).  This means that in the case of a single D string it is
$O(1) = Z_2$.  Thus in this case there are two possible values for the Wilson
line $(\pm 1)$.  The dual type I$^\prime$ description is a single D$0$-brane
stuck to one of the orientifold planes, with the value of the Wilson line
controlling which one it is.

A single D$0$-brane of type I$^\prime$ stuck to an orientifold-plane cannot 
move off the plane into the bulk.  However, a pair of them can do so.  To
understand this, let us consider a pair of wrapped D strings of type I,
coincident in the other dimensions, which carries an O(2) gauge symmetry.
Again, this is T dual to a pair of type I$^\prime$ D0-branes with positions
controlled by the choice of O(2) Wilson line.  The inequivalent choices of
Wilson line are classified by conjugacy classes of the O(2) gauge group.  So we
should recall what they are.  It is important that O(2), unlike its SO(2)
subgroup, is non-Abelian.  Correspondingly, there are conjugacy classes of two
types:
\begin{itemize}
\item The SO(2) subgroup has classes labeled by an angle $\theta$.  Including
the effect of the reflection, inequivalent classes correspond to range $0 \leq
\theta \leq \pi$.  Such a conjugacy class describes a D$0$-brane at $X' =
\theta R'$ in the bulk, together with the mirror image at $X' = (2\pi - \theta)
R'$.  We see that to move into the bulk a second (mirror image) D0-brane had to
be provided.
\item The reflection elements of O(2) all belong to the same conjugacy class.
A representative is the matrix $\left(\begin{array}{cc} 1 & 0\\ 0 &
-1\end{array}\right)$.  This class corresponds to one stuck D$0$-brane on each
O8-plane.
\end{itemize}

\section{Brane Creation}

The solutions of massive type IIA supergravity were investigated by Polchinski
and Witten,\cite{polchinski95a}
who showed that they involve a metric and dilaton that vary in one
direction.  In the context of the type I$^\prime$ theory this means they vary
in all regions for which the number of D$8$-branes to the left and to the right
are unequal.  Thus the only case for which this effect does not occur is the
SO(16) $\times$ SO(16) configuration with D$8$-branes attached to each of the
O8-planes. (This case is closely related to the M theory description of the
$E_8\times E_8$ theory.\cite{horava95})

We can avoid describing the $X'$ dependence of the metric explicitly by using
proper distance $s$ as a coordinate along the interval.  (This requires holding
the other coordinates fixed.)  Then one has $0 \leq s \leq \pi R', R' = 1/R$.
We didn't address the issue earlier, but when we said the interval has length
$\pi R'$ we really did mean its proper distance.  In terms of this coordinate
there is a varying dilaton field, and hence a varying string coupling constant
$g_A (s)$.  Only in regions with half of the D$8$-branes to the left and half to the
right is it constant.

The function $g_A (s)$ was obtained by Polchinski and Witten by solving the
field equations.  A more instructive way of obtaining and understanding the
result uses the brane creation process.  Consider an isolated
D$0$-brane in a region where $g_A(s)$ is constant.  Now suppose the D$0$-brane
crosses a D$8$-brane to enter a region where $g_A(s)$ is varying.  What happens
is that the D$0$-brane emerges on the other side with a fundamental string
stretched between it and the D$8$-brane.  This phenomenon, called Hanany--Witten
effect,\cite{hanany96}  has been derived by a variety of means.\cite{bachas97a} 
It occurs in many different
settings that are related by various duality transformations.  (For example,
two suitably oriented M5-branes can cross to give rise to a stretched
M2-brane.\cite{dealwis97})  The intuitive reason that string creation is required can be
understood as follows.  The original D$0$-brane configuration preserved half
the supersymmetry and was BPS.  Therefore a delicate balance of focus ensured
that it was stable at rest.  When it crosses the D$8$-brane (adiabatically) the
amount of supersymmetry remains unchanged and so it should still be stable at
rest.  To be specific, let us consider the D$8$-brane configuration discussed
earlier with $8 + N$ D$8$-branes on the $X' = 0$ O8-plane and $8 - N$
D$8$-branes on the $X' = \pi R'$ O8-plane.  In this case $N$ fundamental
strings should connect the D$0$-brane to the $X' = 0$ O8-plane.  The BPS
condition implies that the mass of the D$0$-brane should be independent of its
position in the interval.  Recalling that the mass of a type IIA D$0$-brane is
$1/g_A$, we therefore conclude that for this configuration
\begin{equation}
\label{Dmass}
M_{D0} = \frac{1}{g_A (0)} = \frac{1}{g_A(s)} + N T_{F1} s.
\end{equation}
Here $T_{F1} = \frac{1}{2\pi}$ is the tension of a fundamental type IIA string
(in string units).  We therefore see that $g_A(s)$ is the reciprocal of a
linear function whenever $N\not= 0$.  Thus, for $N\not= 0$ it necessarily
develops a pole if $R'$ is too large.

The mass $M_{D0}$ can also be computed in the type I picture in terms of a pair
of wrapped D strings with Wilson line.  The mass is independent of the O(2)
Wilson line, since it is independent of the $X'$ coordinate.  However, it does
depend on the SO(32) Wilson line.  Altogether the mass is a sum of two
contributions:
\begin{equation}
M_{D0} = M_{\rm winding} + M_{\rm Wilson} .
\end{equation}
The winding term contribution is given by simple classical considerations:
\begin{equation}
M_{\rm winding} = 2 \cdot 2\pi R \cdot T_{D1} = \frac{2R}{g}.
\end{equation}
A more careful analysis is required to obtain the Wilson line contribution
\begin{equation}
M_{\rm Wilson} = \frac{N}{4R} .
\end{equation}
Note that this contribution vanishes for large $R$.

We now come to the main point.  There is a special value of $R'$, the one for
which the coupling diverges at the $X' = \pi R'$ orientifold plane.  In this
case
\begin{equation}
\label{vanish}
\frac{1}{g_A (\pi R')} = 0,
\end{equation}
which implies, using eq. (\ref{Dmass}), that
\begin{equation}
M_{D0} =  \frac{N}{2R} = \frac{2R}{g} + \frac{N}{4R},
\end{equation}
and hence that
\begin{equation}
R^2 = gN/8.
\end{equation}
This is precisely the value that we previously asserted gives the symmetry
enhancement SO($16 - 2N$) $\times$ U(1) $\rightarrow E_{9-N}$.  The reason that
there is symmetry enhancement is that there are additional massless vectors
with appropriate quantum numbers.  They arise as the ground states of open
strings connecting the D$8$-branes to a stuck D$0$-brane.\cite{bergman97,bachas97}
This works because
the stuck D$0$-brane is massless in this case, as a consequence of eq. (\ref{vanish}).  
This accounts for all the extra gauge bosons when $N>2$. In the $E_7$ and $E_8$ cases,
there are  additional states attributable to a single bulk D0-brane near $X' = \pi R'$.

\section{Conclusion}

The study of supersymmetric theories has come a long way since Golfand's
pioneering work. I presume that he would be pleased.

\section*{Acknowledgments}
I am grateful to O. Bergman for very helpful discussions. This work was
supported in part by the U.S. Dept. of Energy under Grant No.
DE-FG03-92-ER40701.

\end{document}